\documentclass[lettersize,journal]{IEEEtran}
\usepackage{amsmath,amsfonts}
\usepackage{algorithmic}
\usepackage{array}
\usepackage{ulem}
\usepackage[caption=false,font=normalsize,labelfont=sf,textfont=sf]{subfig}
\usepackage{textcomp}
\usepackage{stfloats}
\usepackage{xurl}
\usepackage{verbatim}
\usepackage{graphicx}
\usepackage{xcolor}
\usepackage{siunitx}
\DeclareSIUnit\torr{Torr}
\usepackage[version=4]{mhchem}
\def\BibTeX{{\rm B\kern-.05em{\sc i\kern-.025em b}\kern-.08em
    T\kern-.1667em\lower.7ex\hbox{E}\kern-.125emX}}
\usepackage{balance}
\usepackage[colorlinks=true, linkcolor=blue, citecolor=blue, urlcolor=blue]{hyperref}

\begin{document}
\title{Versatile Sleeve-and-Bulk Fabrication of Inversely Designed Nanophotonic Structures with High Pattern Transfer Fidelity}

\author{
    Lan Hoang Mai,
    Radheya Sham Sarode,
    William G. Eshbaugh,
    Sulaiman Al Ghadani,
    Tyler Manfre,
    Nazifa Tasnim Arony,
    Joshua M. O. Zide,
    Matthew F. Doty,
    and Edward B. Flagg

\thanks{L. H. Mai, R. Sarode, T. Manfre, and M. F. Doty are with the Quantum Science and Engineering Program, University of Delaware, Newark, DE 19716 USA.}
\thanks{W. Eshbaugh, S. Al Ghadani, and E. B. Flagg are with the Department of Physics and Astronomy, West Virginia University, Morgantown, WV 26506 USA.}
\thanks{N. T. Arony and J. M. O. Zide are with the Department of Materials Science and Engineering, University of Delaware, Newark, DE 19716 USA.}

}

\maketitle

\begin{abstract}
Inverse design is a powerful computational tool for generating nanophotonic devices that should outperform their traditionally designed counterparts. However, translating these theoretical gains into physical devices is challenging because the irregular and non-intuitive features that make inversely designed structures so effective also make them exceptionally difficult to fabricate, typically necessitating substantial process optimization unique to each device design. We describe a versatile two-step nanofabrication process that achieves pattern transfer fidelity in excess of \qty{90}{\percent} for each of three distinct inverse device designs with no device-specific process optimization.
\end{abstract}


\section{Introduction}
\IEEEPARstart{R}{ealizing} practical, scalable quantum photonic technologies requires the seamless integration of quantum emitters into photonic nanostructures. These applications demand devices that operate near theoretical limits, requiring high coupling efficiencies~\cite{faraon_efficient_2007,senellart_high-performance_2017}, ultra-small mode volumes~\cite{akahane_high-q_2003,lodahl_interfacing_2015}, and low-loss waveguides~\cite{uppu_quantum-dot-based_2021,moody_2022_2022,chanana_ultra-low_2022}. Historically, the design of these components has been guided by physical intuition, and while this forward-engineering approach has yielded numerous high-performance devices---such as bullseye cavities~\cite{wang_-demand_2019, davanco_circular_2011, liu_solid-state_2019, rickert_high_2024}, grating couplers~\cite{liu_high-efficiency_2024,carroll_broad_2013}, and photonic crystals~\cite{akahane_high-q_2003, yoshie_vacuum_2004}---relying strictly on human intuition and simple models inherently limits the explorable design space.

\begin{figure}[!b]
    \centering
    \includegraphics[width=\columnwidth]{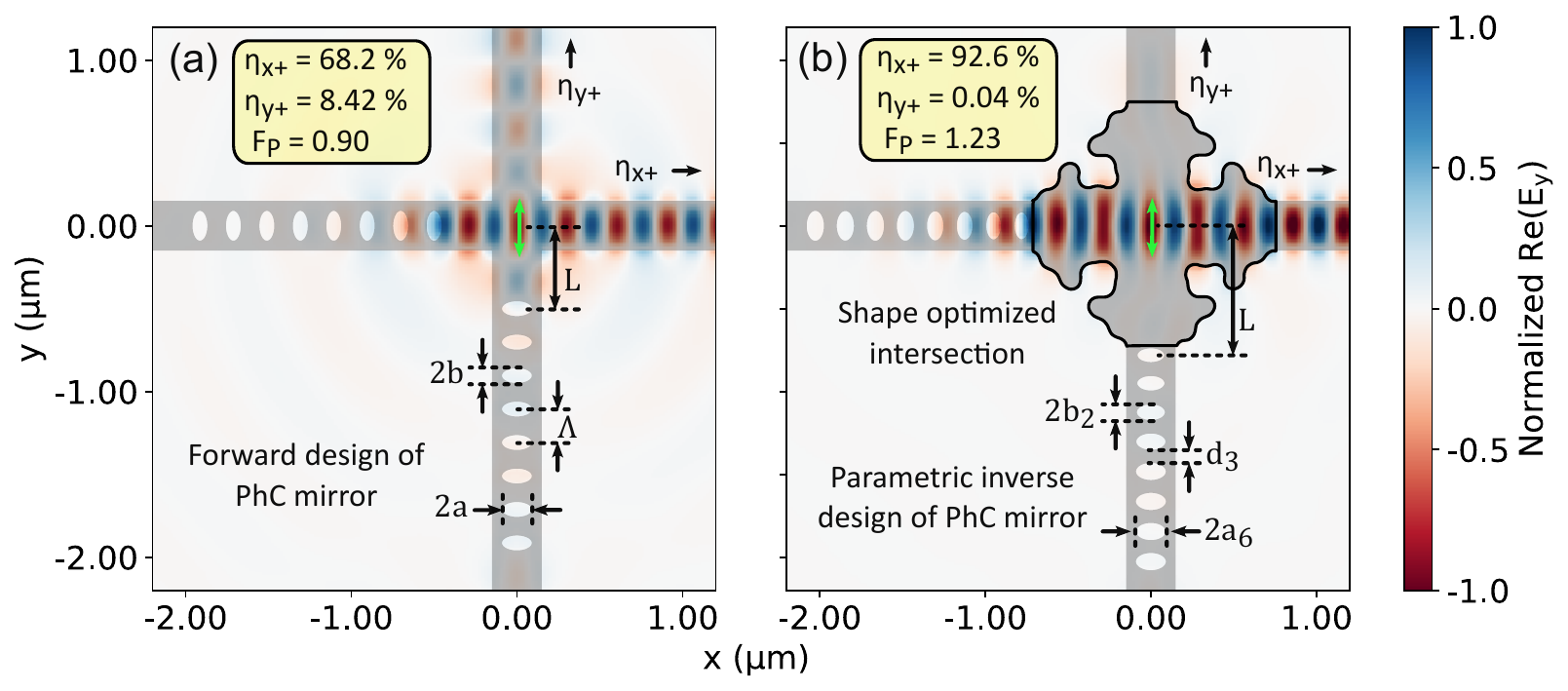}
    \caption{Conventional versus inverse design of a waveguide crossing (WGX) for polarization-selective coupling of electric dipole radiation into separate waveguides. (a) Simple WGX with conventionally forward-designed photonic crystal mirrors obtained via a grid sweep over a small number of parameters: the pitch $\Lambda$, and the dipole-to-mirror distance $L$. The coupling efficiency to the desired fundamental waveguide mode ($\eta_{x+}$) is \qty{68.2}{\percent}, the Purcell factor ($F_P$) is 0.9, and crosstalk ($\eta_{y+}$) is \qty{8.42}{\percent}. (b) An inversely designed WGX featuring a shape-optimized intersection and photonic crystal holes, where the semi-major axis $a_i$, semi-minor axis $b_i$, spacing $d_i$ of the $i$-th hole, along with the distance $L$ between the first hole and dipole, are optimized through a gradient descent algorithm. The optimized device has a coupling efficiency to the target waveguide mode of \qty{92.6}{\percent},  a Purcell factor of 1.23, and crosstalk of \qty{0.04}{\percent}---a significant improvement over the conventional counterpart---demonstrating the powerful utility of inverse design.}
    \label{fig:wgx_design}
\end{figure}

To overcome this constraint, inverse design (ID) has rapidly gained traction across all facets of integrated quantum photonics in recent years~\cite{molesky_inverse_2018}, including photon sources~\cite{wambold_adjoint-optimized_2021,melo_multiobjective_2022,melo_inverse_2023}, coupling interfaces~\cite{lu_objective-first_2012,michaels_inverse_2018,huang_compact_2025,dory_inverse-designed_2019,hansen_efficient_2023}, metasurfaces~\cite{fan_freeform_2020,yang_exploring_2025, cai_inverse_2020}, power splitters~\cite{tahersima_deep_2019, song_ultracompact_2024,hansen_inverse_2024}, wavelength demultiplexers~\cite{piggott_inverse_2015, vercruysse_analytical_2019}, and mode converters~\cite{lu_objective-first_2012, vercruysse_analytical_2019}. In contrast to conventional approaches often using semi-analytical parameter sweeps, ID frames the design process as a computational optimization problem, where it automatically refines device geometries by minimizing an objective function---which mathematically encapsulates the desired performance metrics---through gradient-based optimization. Typically, ID leverages the adjoint method, which enables efficient gradient calculations using the electromagnetic fields obtained via finite-difference time- or frequency-domain simulations, independent of the number of design parameters~\cite{niederberger_sensitivity_2014,giles_introduction_2000,minkov_inverse_2020}. This decoupling of computational cost from structural complexity allows algorithms to explore a vastly expanded parameter space. ID spans a broad spectrum of techniques, ranging from basic parameterization of etched feature dimensions~\cite{asano_optimization_2018, vij_inverse_2024} and shape optimization~\cite{lalau-keraly_adjoint_2013}, to full topology optimization~\cite{jensen_topology_2011, christiansen_inverse_2021} exploring the entire continuous space of dielectric permittivity within a given footprint---all of which can produce devices with performance matching (or exceeding) their conventionally designed counterparts~\cite{molesky_inverse_2018,piggott_inverse-designed_2020,carfagno_inverse_2023}. Figure~\ref{fig:wgx_design} compares a forward-designed device to its inverse-designed counterpart; the coupling efficiency improves from approximately \qty{68}{\percent} (forward-design) to approximately \qty{93}{\percent} (inverse-design). Beyond this simple example, exploring even more of the design space through topology optimization yields even higher theoretical performance~\cite{melo_inverse_2023}. However, translating these theoretical gains into physical devices introduces a significant challenge: fabrication. The irregular and non-intuitive features that make ID structures so effective often make them exceptionally difficult to manufacture, and can require tailoring the design process to incorporate fabrication constraints~\cite{vercruysse_analytical_2019,michaels_leveraging_2018,hammond_photonic_2021,piggott_inverse-designed_2020, khoram_controlling_2020} along with more robust fabrication techniques~\cite{carfagno_sleeve_2022}.

Here we demonstrate a versatile sleeve-and-bulk process that allows us to fabricate multiple inversely designed devices with high fidelity to the design geometry with no fabrication process optimization. We first describe the fabrication strategy and implementation details. We then describe our methods for quantifying the pattern transfer fidelity (PTF). Finally, we show that a single fabrication run using this method, with no process optimization, can render three distinct ID devices with greater than \qty{90}{\percent} PTF for each device.  

\section{Fabrication Strategy} 
\begin{figure}[!t]
    \centering
    \includegraphics[width=\columnwidth]{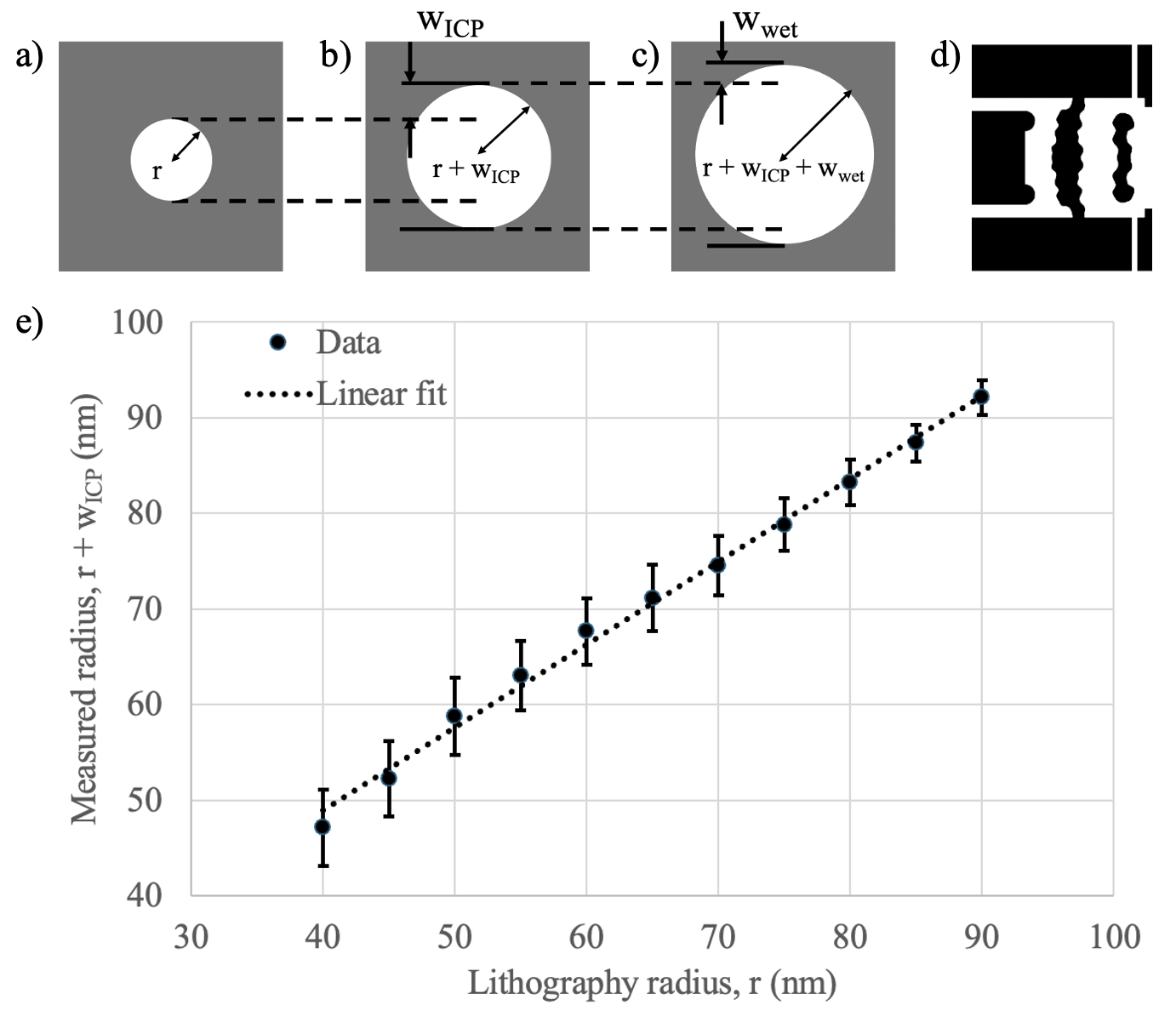}
    \caption{(a) Example feature as defined by lithography, with radius r. (b) Etch bias introduced by dry etch (w$_{\text{ICP}}$) enlarges the hole to a total radius of r + w$_{\text{ICP}}$. (c) Etch bias introduced by hydrofluoric undercut and digital clean (w$_{\text{wet}}$) further expands the hole to a total radius of r + w$_{\text{ICP}}$ + w$_{\text{wet}}$. (d) Continuous variation of feature size characteristic of topologically optimized ID devices. (e) Calibration data showing experimentally measured hole radius (r + w$_{\text{ICP}}$) after ICP etch as a function of the feature radius written by electron beam lithography (lithography radius, r) for a 2 minute 30 second etch. The calibration data is fitted with a linear model, yielding $y = 0.8657 x + 14.2$ where $y$ is the measured radius (r + w$_\text{ICP}$) and $x$ is the lithography radius, r.}
    \label{fig:etch_bias}
\end{figure}

Inversely designed photonic devices---especially topology-optimized designs \cite{hammond_designing_2019, yesilyurt_efficient_2021}---often have features on multiple length scales. This creates a substantial fabrication challenge that originates in a phenomenon known as etch bias. As shown in Fig.~\ref{fig:etch_bias}(a-c), etch bias---the over-etching of device features---is introduced at both the inductively coupled plasma (ICP) etch (Fig.~\ref{fig:etch_bias}(b)) and subsequent wet etch (Fig.~\ref{fig:etch_bias}(c)) fabrication steps. If this etch bias were of constant amplitude, it would be a simple matter to compensate for it by reducing the size of the features defined by e-beam lithography. However, the etch bias that arises from ICP is feature size dependent because the feature size determines the rates at which fresh etchants enter the feature and waste products are removed. This poses a particularly large challenge for topologically-optimized structures that vary continuously between ``large" and ``small" features, as schematically depicted in Fig.~\ref{fig:etch_bias}(d). In order to compensate for these feature-size--dependent etch rates and achieve the vertical sidewall profiles required for minimal scattering, inversely designed devices typically require either multiple etch recipes calibrated for different length scales or feature-specific edge biases in the design. Thus, optimizing an etch recipe for each device design typically requires substantial time, effort, and material. 

The nanofabrication process we describe here solves the problem of non-uniform etch biases with two complementary methods: (1) pre-correction of the mask file for small closed features (e.g., holes, slots) whose narrowest dimension is less than \qty{220}{\nano\meter} and (2) sleeve-and-bulk fabrication of large features (i.e., with narrowest dimension $\geq$~\qty{220}{\nano\meter}). To determine the pre-correction of the mask, we generated a large calibration data set derived from SEM images of photonic crystals with different hole sizes and created a look-up table that reports the ICP etch bias (w$_{\text{ICP}}$) as a function of both the target feature area and etching time. See Fig.~\ref{fig:etch_bias}(e) for example calibration data for etch times of 2 minutes and 30 seconds. We pre-correct the mask file by reducing the size of every closed feature (circle, ellipse, etc.) by the value of w$_{\text{ICP}}$ appropriate to the area of that feature and the expected etch time. 

For ``large" features, we overcome the challenge of non-uniform etch bias by using our sleeve-and-bulk method \cite{Carfagno_2023}, which divides the fabrication into two steps: sleeve etch and bulk etch. This method is inspired by the sleeve-and-bulk exposure strategy for electron beam lithography where the device boundaries, or ``sleeves", are written with a small beam size to maximize accuracy while the inner areas, or ``bulk", are written with a large beam size to minimize the exposure time. For a family of inversely designed devices, a fixed sleeve width can be applied to define the perimeter of all large features because the sleeve width does not depend on the shape or feature variation between devices. That means a single optimized dry-etch recipe can consistently transfer the sleeve pattern into the substrate with vertical sidewall profiles and desired etch depth. A second lithography and etch step removes the remaining bulk. The conditions for the bulk etch are more forgiving because it only needs to etch the inner feature area. The vertical sidewall profile, critical to photonic device performance, is protected by the resist during the bulk etch. The closed features whose pre-correction is described above can be fabricated in either the sleeve or bulk steps, whichever is more appropriate for the area of those closed features.

\begin{figure*}[!h]
    \centering
    \includegraphics[width=2\columnwidth]{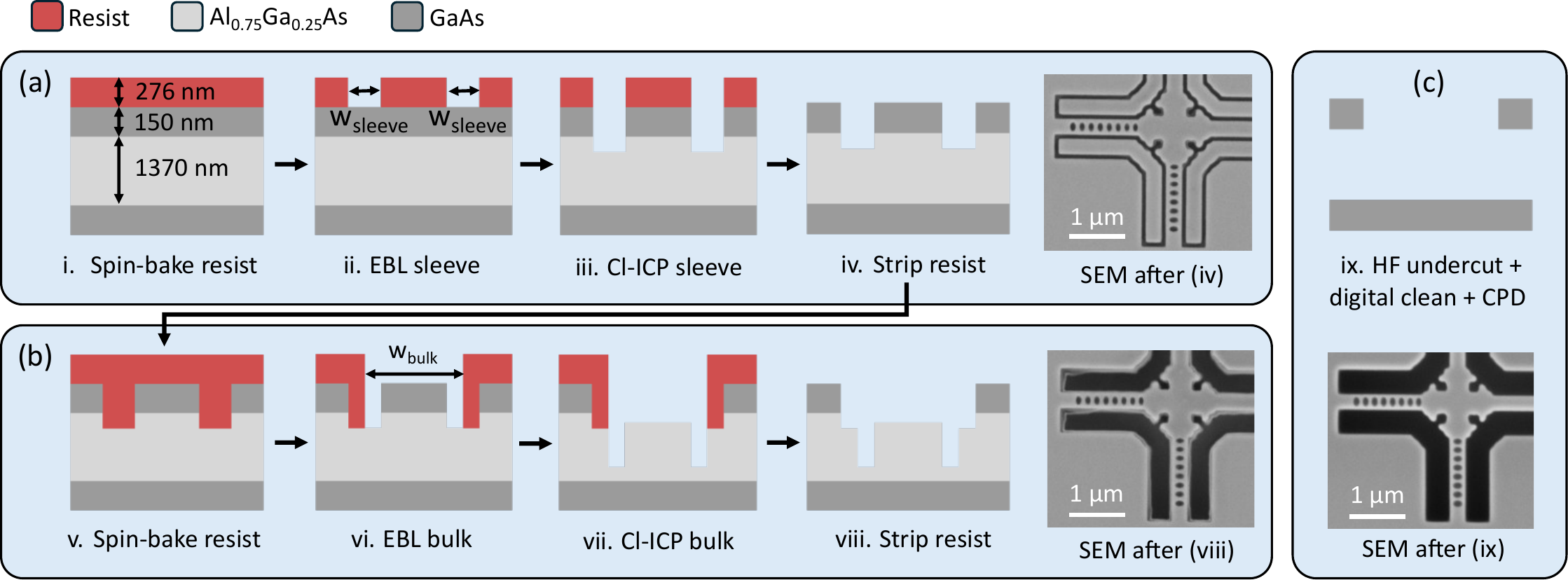}
    \caption{
        Sleeve and bulk process diagram. (a) Sleeve lithography and etch steps and corresponding SEM image, where w$_{\text{sleeve}}$ is the width of the sleeve feature. (b) Bulk lithography and etch steps and corresponding SEM image, where w$_{\text{bulk}}$ is the width of the bulk feature. (c) HF sacrificial layer removal, digital cleaning, and CPD with final SEM image of a fully suspended device.
    }
    \label{fig:process_diagram}
\end{figure*}

\section{Fabrication Details}
The inversely designed polarization demultiplexer shown in Fig.~\ref{fig:wgx_design}(b) was optimized using Tidy3D's finite-difference time-domain solver \cite{flagg_shape-optimized_2026}. We use it as an example to describe how we employ this versatile sleeve and bulk-based fabrication strategy. The fabrication process along with scanning electron micrographs after the sleeve, bulk, and undercut etches are summarized in Fig.~\ref{fig:process_diagram}. The device is fabricated on a \qty{150}{\nano\meter} \ce{GaAs} membrane grown on top of a \qty{1370}{\nano\meter} \ce{Al_{0.75}Ga_{0.25}As} sacrificial layer. Both materials are grown by molecular beam epitaxy on a (100) \ce{GaAs} substrate (Fig.~\ref{fig:process_diagram}(a)). 

Prior to fabrication, the sample is sonicated in N-Methylpyrrolidone (NMP) and rinsed with isopropyl alcohol (IPA) to remove organic residues. For the sleeve fabrication, electron beam resist AR-P6200.09 is spun on the sample at \qty{3000}{rpm} for 1 minute, then baked at \qty{170}{\celsius} for 5 minutes to evaporate the resist solvent. The resist is exposed with electron beam lithography (EBL) using a Raith EBPG 5200 at \qty{200}{\micro\coulomb\per\centi\meter\squared} at a beam current of \qty{1}{\nano\ampere}. Dual-pass exposure (multi-pass N = 2) is used to reduce the sleeve's line-edge roughness. The written sleeve width w$_\text{sleeve}$ is chosen to be \qty{62}{\nano\meter} \cite{Carfagno_2023}. Since the mirror ellipses of the design are within the length scale where the sleeve etch recipe can produce vertical sidewalls, they are exposed and etched simultaneously with the sleeves. To account for feature enlargement during dry etching, the ellipses are biased smaller than designed following the calibration data in Fig.~\ref{fig:etch_bias}(e). The resist is then developed with AR 600-546 for 2 minutes, rinsed with IPA, blow dried with \ce{N_{2}} before an \ce{O_{2}} plasma descum for 1 minute. 

To transfer the sleeve patterns, the sample is etched for 2 minutes 40 seconds in a chlorine inductively coupled plasma (Cl-ICP) etcher (PlasmaTherm Apex SLR) with gas flow rates of \qty{15}{sccm} \ce{BCl_{3}} and \qty{10}{sccm} \ce{Ar} at \qty{6}{\milli\torr} of pressure. The ICP coil power and bias power are set to \qty{500}{\watt} and \qty{25}{\watt}, respectively. To strip the resist, the sample is soaked in NMP at \qty{80}{\celsius} for 2 hours before being sonicated in the same NMP beaker for 15 minutes. The bulk pattern transfer uses the same lithography, development, etching, and cleaning steps as those used for the sleeve transfer.

The device is suspended by etching the \ce{Al_{0.75}Ga_{0.25}As} sacrificial layer in hydrofluoric acid (HF \qty{5}{\percent}) (Fig.~\ref{fig:process_diagram}(c)). HF etch residues (such as aluminum fluoride \ce{AlF_{3}}, aluminum hydroxide \ce{Al(OH)_{3}}) and any hardened e-beam resist due to Cl-ICP are then removed with a digital cleaning procedure \cite{Carfagno_2023, Midolo_soft-mask_2015}: the sample is soaked in hydrogen peroxide (\ce{H2O2} \qty{30}{\percent}) for 1 minute, rinsed with de-ionized water for 1 minute, soaked in potassium hydroxide (\ce{KOH} \qty{22.5}{\percent}) for 1 minute, and rinsed with de-ionized water for 1 minute. Finally, the sample was dried by \ce{CO_{2}} critical point drying (CPD). 

All SEM images were taken with a Zeiss Merlin HR-SEM with \qty{5}{\kilo\volt} accelerating voltage, \qty{25}{\pico\ampere} probe current, and \qty{3.6}{\milli\meter} to \qty{3.7}{\milli\meter} of working distance. SEM drift was compensated using the ``Drift Compensated Frame Integration" mode with the number of frames to integrate set to N = 10. Each image has a resolution of $4096\times3072$ pixels, and the scan speed is set to 4 so that each image is acquired in 1.8 minutes. All SEM images were taken at the same brightness and contrast.

\section{Quantifying Pattern Transfer Fidelity}
Fabrication fidelity quantifies how closely a fabricated device geometry matches the intended design. Existing fabrication fidelity methods are often developed for specific types of devices with simple and/or repetitive periodic geometry. For example, the method implemented in~\cite{Carfagno_2023_IOP} works by binarizing the SEM image of a photonic crystal and extracting feature-level parameters like hole centroid position, size, and shape and comparing them with the feature-level parameters extracted from the original GDS mask file. This approach is well suited for photonic crystal structures because the geometry consists of repeated, identifiable features that can be compared to an ideal lattice from a GDS mask. A more general method used in semiconductor metrology is edge-placement-error (EPE) \cite{sato_edge_2019}, which computes the distance between an SEM-extracted contour and an ideal GDS-extracted contour. While EPE provides detailed information about the edge displacement at all locations, it does not define a simple device-level fidelity score that can be applied to arbitrary ID patterns. To overcome these limitations, we define and describe a fabrication fidelity metric called pattern transfer fidelity (PTF). In addition to quantifying overall fabrication fidelity, PTF can also be used as an intermediate step in device fabrication and characterization workflow to quantify the performance of individual fabrication steps. It also provides a computationally cheap tool for estimating how a fabricated photonic device might perform.

PTF is calculated by converting both the SEM of a fabricated device and the GDS describing that device's target structure to binary images in which each pixel is either material (white, 1) or void (black, 0). We define the PTF using an intersection-over-union (IoU) metric that quantifies the pixel-wise overlap between the GDS and SEM binary images. IoU is already widely used to evaluate fabrication fidelity in nanophotonic devices \cite{azimi_semu-net_2024, azimi_gen-fab_2026} and defect metrology in semiconductor manufacturing \cite{dehaerne_scanning_2025}. Before a PTF can be computed, however, the SEM image file must be calibrated, denoised, and converted from grayscale to black and white. We first describe how we prepare the SEM images, then describe how we compute the IoU and PTF.

\subsection{SEM image preparation}
\begin{figure*}[b!]
    \centering
    \includegraphics[width=2\columnwidth]{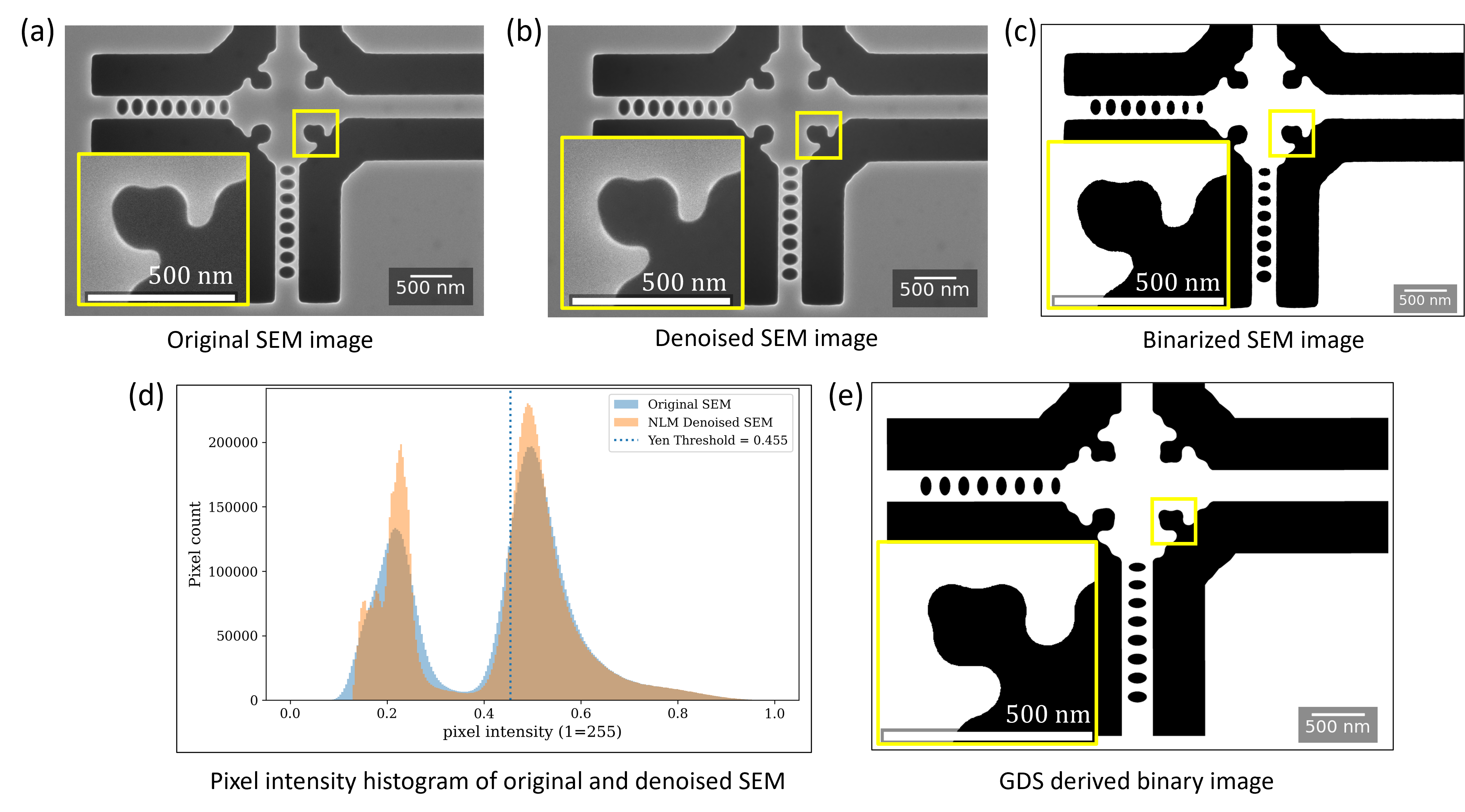}
        \caption{(a) Original SEM image. (b) SEM image denoised with NLM. (c) Binarized SEM image using Yen thresholding. (d) Histogram of pixel intensity for both the original and denoised SEM images with the Yen threshold (0.455 pixel intensity) indicated. (e) GDS binary image after alignment optimization (translation, rotation). All inset scale bars are \qty{500}{\nano\meter} in length.}
    \label{fig:sem_binarization_ptf}
\end{figure*}

Any meaningful physical measurement, like the PTF, depends on the physical pixel size extracted from the SEM image metadata. We used an established calibration routine~\cite{copeland_traceable_2024, madison_total_2023} to quantify the SEM scale factor and scan-field distortion present in our system and found a systematic error of approximately \qty{1.9}{\percent} between the scale factor provided by the manufacturer and that returned by our calibration routine. This is within the precision specified by the manufacturer ($\pm~$\qty{2}{\percent}). We use the calibrated scale factor in all of our image analyses.

We next denoise the SEM image to reduce pixel intensity noise while preserving the location of the material-air edges. We tested two edge preserving methods: bilateral filtering \cite{710815} and non-local means denoising (NLM) \cite{1467423}. The bilateral filter preserves edges by averaging neighboring pixels only when they are spatially close and have similar intensity to the target pixel.\footnote{We use a 75-pixel neighbor diameter for the bilateral filter.} NLM first selects a small template patch around a target pixel and identifies all similar template patches within a given search window. NLM then replaces the target pixel with a weighted average of the centers of the similar template patches, with the weights determined by (a) how similar each template patch is to the template patch around the target pixel and (b) the filter strength specified by the user.\footnote{We use the optimal NLM denoising parameters for best edge preservation as specified in OpenCV documents: a 15$\times$15 pixel template patch around the target pixel, a search for similar template patches in a 43$\times$43 pixel search window, and filter strength between 5 and 12.} We found that bilateral filtering and NLM produced similar values of PTF. An example of an original and NLM-denoised SEM image are shown in Figs.~\ref{fig:sem_binarization_ptf}(a) and (b), respectively. The effect this denoising has on the histogram of SEM image pixel intensities is shown in Fig.~\ref{fig:sem_binarization_ptf}(d). In such histograms, the pixel intensity is normalized by the maximum possible value (in our case 255) so that pixel intensities are reported on a scale of 0 (black) to 1 (white). 

Finally, we use the Yen thresholding method to convert the denoised SEM image into a binary image. Yen thresholding is a histogram-based thresholding method that selects a threshold by maximizing a correlation-based criterion between separated pixel classes \cite{366472}.  All pixels with intensity below the threshold are set to 0 (black) and all pixels above the threshold are set to 1 (white). Yen thresholding does not require fitting the histogram to an assumed model, and we found its maximum-correlation histogram criterion is less affected by histogram noise (see Fig.~\ref{fig:sem_binarization_ptf}(d)) than other thresholding methods we explored. For example, residual noise and intermediate greyscale pixels can make valley-based \cite{NG20061644} or variance-based thresholding \cite{4310076} methods less reliable. See Fig.~\ref{fig:sem_binarization_ptf}(c) for an example of an SEM image binarized using a Yen threshold of 0.455 pixel intensity, as indicated by the dashed line in Fig.~\ref{fig:sem_binarization_ptf}(d). 

\subsection{Calculating IoU and PTF}
The binarized SEM image, $S$ (e.g., Fig.~\ref{fig:sem_binarization_ptf}(c)) can now be compared to a binarized GDS image. The binarized GDS image, $G$ (e.g., Fig.~\ref{fig:sem_binarization_ptf}(e)) is simply generated by pixelating the GDS mask file with the same physical pixel size (nm/pixel) as the calibration-corrected SEM image and extracting an image with the same pixel dimensions as the SEM binary image. 
Mathematically, these two images are defined by pixel values $G_{ij}$ and $S_{ij}$, where $(i,j)$ are the pixel indices and each pixel has a value of either 1 (material) or 0 (void).
The standard material IoU, which measures the overlap of the material regions between $G$ and $S$ is defined as
\begin{equation}
\label{eq:material_iou}
\text{IoU}_{\text{material}} = \frac{|\mathcal{G} \cap \mathcal{S}|}{|\mathcal{G} \cup \mathcal{S}|}
\end{equation}
where
\begin{equation}
\begin{aligned}
\label{eq:define_set}
\mathcal{G}=\{(i,j)\in\Lambda:G_{ij}=1\}, 
\\  
\mathcal{S}=\{(i,j)\in\Lambda:S_{ij}=1\}.
\end{aligned}
\end{equation}
Here $\Lambda$ is restricted to the set of pixel indices $(i,j)$ in the range of interest for the IoU calculation, and $\mathcal{G}$ and $\mathcal{S}$ are the subsets of pixels for which ${G}_{ij},{S}_{ij} =1$, i.e., the regions corresponding to material. 

We can also write the material IoU in terms of the confusion matrix elements TP, TN, FP, and FN. As we will show below, these confusion matrix elements allow us to directly visualize the success or failure of the fabrication process at each pixel. TP (true positive) represents regions where material is present in the SEM image and the GDS image says material should be there. FP (false positive) represents material which appears in the SEM image but not the GDS image. FN (false negative) represents material missing in the SEM but present in the GDS. And TN (true negative) represents void/air in both the SEM and GDS images. TP, TN, FP, and FN are defined by simple set operations on $G_{ij}$ and $S_{ij}$:
\begin{equation}\label{eq:confusion_matrix_eqs}
\begin{aligned}
\text{TP} &= \sum_{(i,j) \in \Lambda} G_{ij}S_{ij}, \  & \text{FN} &= \sum_{(i,j) \in \Lambda} G_{ij}(1 - S_{ij}), \\
\text{FP} &= \sum_{(i,j) \in \Lambda} (1 - G_{ij})S_{ij}, \  & \text{TN} &= \sum_{(i,j) \in \Lambda} (1 - G_{ij})(1 - S_{ij}).
\end{aligned}
\end{equation}
which yields the alternative definition of the material IoU
\begin{equation}\label{eq:iou_extended}
\text{IoU}_{\text{material}} = \frac{|\mathcal{G} \cap \mathcal{S}|}{|\mathcal{G} \cup \mathcal{S}|} = \frac{\text{TP}}{\text{TP} + \text{FP} + \text{FN}}. 
\end{equation}

The material IoU defined by Eqn.~(\ref{eq:iou_extended}) quantifies locations where the fabrication process correctly left material in place. However, correctly removing material to create air voids is equally important for nanophotonic devices. We can similarly use the confusion matrix elements given by Eqn.~(\ref{eq:confusion_matrix_eqs}) to compute an air IoU, which is given by
\begin{equation}\label{eq:air_ioU}
\text{IoU}_{\text{air}} = \frac{\text{TN}}{\text{TN} + \text{FP} + \text{FN}}.
\end{equation}

The final PTF is defined by taking the average of the material and air IoU terms defined in Eqns.~(\ref{eq:iou_extended}) and (\ref{eq:air_ioU}), respectively:
\begin{equation}\label{eq:balanced_iou_parentheses}
\text{IoU}_{\text{balanced}} = \frac{1}{2} \left( \frac{\text{TP}}{\text{TP} + \text{FP} + \text{FN}} + \frac{\text{TN}}{\text{TN} + \text{FP} + \text{FN}} \right)
\end{equation}

We have now defined the PTF mathematically, but one practical challenge must still be addressed: to obtain an accurate PTF, the GDS and SEM binary images must be aligned via rotation and translation. To accomplish this we perform a local grid search that optimizes the PTF. First, the GDS binary image is coarsely aligned to the SEM binary image by manually translating the center of the range of interest and rotating with respect to the GDS origin. The grid search then applies small perturbations around this manual initial alignment by testing all unique combinations of $x$- and $y$-translation offsets from $-$\qty{5}{\nano\meter} to $+$\qty{5}{\nano\meter} and rotation offsets from $-$\ang{0.7} to $+$\ang{0.7} with $x$- and $y$-translation offset step size of \qty{0.15} {\nano\meter} and rotation offset step size of \ang{0.1}. A total of 25,125 unique parameter combinations are tested. In each case, a new binarized GDS image is extracted from the mask file using the coordinate origin and rotation angle parameters and the PTF is calculated using Eqn.~(\ref{eq:balanced_iou_parentheses}). Figure \ref{fig:sem_binarization_ptf}(e) shows the optimally aligned GDS binary image. We use the full SEM image during the GDS-SEM alignment optimization because it provides more structural information and makes the alignment optimization less sensitive to local noise. After the alignment optimization is complete, the final PTF is calculated only on predefined analysis areas (see yellow boxes in Fig.~\ref{fig:Confusion_map}). This avoids including large areas of either material or air/void that would artificially increase the PTF value. 

\section{Results and Discussion}
\begin{figure*}[t]
    \centering
    \includegraphics[width=2\columnwidth]{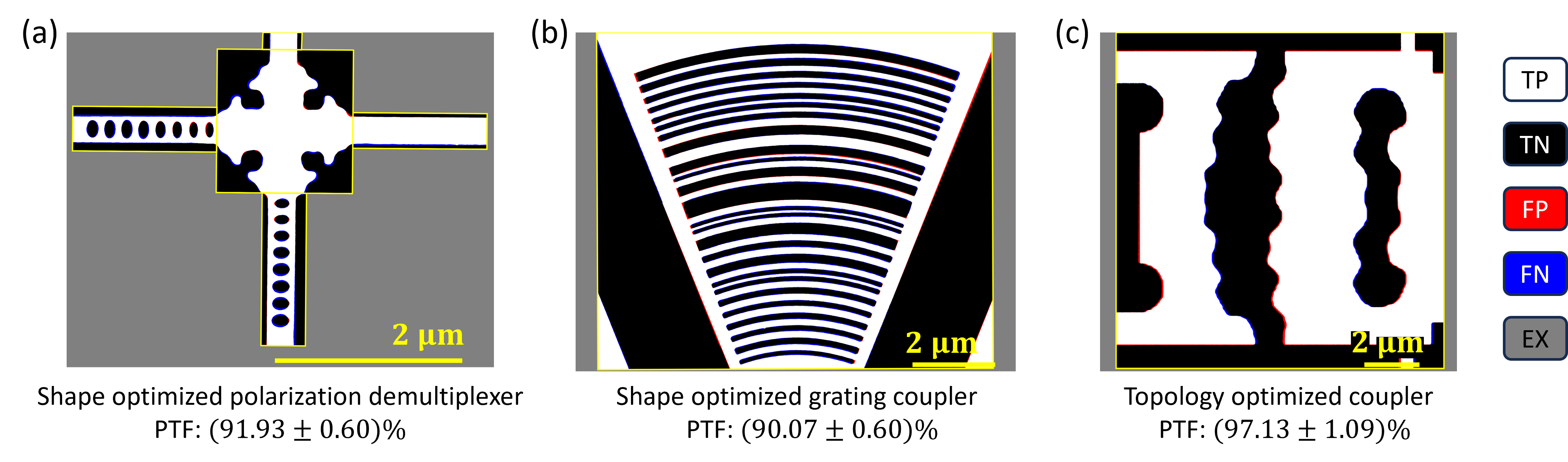}
        \caption{PTF calculated on three different inversely designed devices. (a) Shape-optimized polarization demultiplexer with photonic crystal mirrors on bottom and left waveguides. PTF: \qty{91.93 \pm 0.60}{\percent}. (b) Shape-optimized grating coupler. PTF: \qty{90.07 \pm 0.60}{\percent}.
        (c) Topology-optimized coupler. PTF: \qty{97.13 \pm 1.09}{\percent}. TP: True positive (correctly matched material). TN: True negative (correctly matched void/background pixels). FP: False positive (extra material detected in the SEM but absent in the GDS). FN: False negative (material present in the GDS but missing in the SEM). EX: Pixels excluded from the PTF calculation. All scale bars are \qty{2}{\micro\meter} in length.}
    \label{fig:Confusion_map}
\end{figure*}

Figure \ref{fig:Confusion_map} presents a representative confusion map image and the average PTF results for three distinct ID device types, two designed with shape optimization and one designed with topology optimization. All three devices were fabricated on a single chip in a single process run using global mask pre-correction and the two-step sleeve-and-bulk fabrication process, with no device-specific process optimization. In each case the average PTF was computed from analysis of five unique instances of that device. The results show that this method achieves an average PTF in excess of \qty{90}{\percent} for all three device designs, despite their significant differences. Moreover, in each case the standard deviation in the PTF for multiple device instances is less than \qty{1.1}{\percent}, demonstrating the consistency of this fabrication method across repeated instances. 

Figure \ref{fig:Confusion_map} provides a clear illustration of how the confusion matrix elements allow us to visualize the success or failure of the fabrication process at each pixel. Specifically, the white (TP) and black (TN) coloring shows successful fabrication of material or void at the target location. Red (FP) indicates areas where the fabrication process failed to remove material (e.g., under-etching) and blue (FN) indicates areas where the fabrication process removed material that should have remained (e.g., over-etching). In the specific case of Fig.~\ref{fig:Confusion_map}, we can see that the red (FP) and blue (FN) regions are small and essentially randomly distributed along boundaries. This indicates the absence of systematic error. More generally, this confusion matrix PTF method provides immediate feedback that could be used to identify and eliminate systematic error when fabrication processes are being optimized. 

The uncertainties reported in Fig.~\ref{fig:Confusion_map} are the standard deviations of the PTFs calculated from the five distinct instances of each device. Another possible source of uncertainty is the choice of SEM image denoising protocol. To assess this we computed the PTF for the inversely designed polarization demultiplexer (Fig.~\ref{fig:wgx_design}(b)) using both bilateral and NLM denoising. When we used bilateral-filter denoising in conjunction with Yen thresholding we obtained a PTF of \qty{93.20 \pm 0.18}{\percent}. When using NLM-denoising with Yen thresholding we obtained a PTF of \qty{93.35 \pm 0.27}{\percent}. These results are identical within the uncertainty, demonstrating that this choice of denoising protocol did not alter the results. Similarly, we assessed the sensitivity of the PTF calculation to uncertainty in the Yen-threshold--defined material edge position. We first used an image-gradient--based edge analysis of the SEM images to estimate our confidence in the material-to-air edge position, obtaining an uncertainty of \qty{\pm2}{\nano\meter}. Edge displacement of \qty{\pm2}{\nano\meter} corresponds to a change in the Yen threshold of approximately \(T_{\mathrm{Yen}} \pm 0.01\). We therefore repeated the PTF calculation for five instances of the inversely designed polarization demultiplexer using \(T_{\mathrm{Yen}} \pm 0.01\). We obtain PTF values of \qty{91.74\pm0.62}{\percent} for a threshold value of \(T_{\mathrm{Yen}} + 0.01\) (edge displaced by \qty{-2}{\nano\meter}) and PTF values of \qty{92.08\pm0.60}{\percent} for a threshold value of \(T_{\mathrm{Yen}} - 0.01\) (edge displaced by $+$\qty{2}{\nano\meter}). The change in mean PTF caused by these edge perturbations is smaller than the device-to-device standard deviation, suggesting that the Yen threshold does not affect the PTF uncertainty.

\section{Conclusions}
We demonstrated a novel nanofabrication process that combines (a) mask pre-correction to overcome feature-size--dependent etch bias with (b) sleeve-and-bulk fabrication to render large features with arbitrary topologies. We then developed a new Pattern Transfer Fidelity (PTF) metric and used it to quantify the fidelity with which this nanofabrication process realizes three distinct inversely designed nanophotonic devices. We find that this nanofabrication method achieved PTF greater than \qty{90}{\percent} for all three devices in a single process run with no device-specific optimization. Moreover, the PTF method provides important information for process optimization should that be necessary. Taken together, these results provide a powerful and flexible nanofabrication method that overcomes a significant practical challenge to realizing the predicted advantages of inversely designed devices. 

\bibliographystyle{IEEEtran}
\bibliography{reference,references_Ned_v4}

\begin{IEEEbiographynophoto}{Lan Hoang Mai}
received the B.S. degree in Math and Physics from McDaniel College in 2022. He is currently pursuing the Ph.D. degree in Quantum Science and Engineering at the University of Delaware, Newark, DE. 

From 2022 to 2023, he was a Graduate Teaching Assistant with the Quantum Science and Engineering Program at the University of Delaware, where he has been a Graduate Research Assistant since 2023. His research interests include III-V photonic devices for single-photon and photon cluster state generation. He is the recipient of the Unidel Distinguished Scholars Award from the University of Delaware.

\end{IEEEbiographynophoto}

\begin{IEEEbiographynophoto}{Radheya Sham Sarode}
received the B.S. degree in Computer Science and Engineering from Dr. D. Y. Patil Institute of Technology, Pune, India, in 2021, and the M.S. degree in Quantum Science and Engineering from the University of Delaware, Newark, DE, USA in 2025. He is currently pursuing the Ph.D. degree in Quantum Science and Engineering at the University of Delaware.

Since June 2025, he has been a Graduate Research Assistant in the Quantum Science and Engineering Program at the University of Delaware. His research interests include inverse design of nanophotonic devices, rare-earth-based light–matter interfaces for quantum photonic technologies, and the characterization of nanophotonic devices, with a particular focus on photonic crystal cavities and waveguides.

\end{IEEEbiographynophoto}

\begin{IEEEbiographynophoto}{William G. Eshbaugh}
received the B.S. degree in applied physics from Ohio University in 2020 and the M.S. degree in physics from West Virginia University in 2023. He is currently pursuing the Ph.D. degree in physics at West Virginia University, Morgantown, WV.

From 2020 to 2023, he was a Graduate Teaching Assistant with the Department of Physics and Astronomy, West Virginia University, where he has been a Graduate Research Assistant since 2023. His research interests include quantum optics, nanophotonics, and nanoscale device design and fabrication, with a particular focus on quantum dots.

Mr. Eshbaugh was a recipient of the Outstanding Graduate Teaching Assistant Award from the West Virginia University Department of Physics and Astronomy in 2023.

\end{IEEEbiographynophoto}

\begin{IEEEbiographynophoto}{Sulaiman Al Ghadani}
received the B.S. degree in physics from West Virginia University in 2022. He is currently pursuing the Ph.D. degree in physics at West Virginia University, Morgantown, WV.

From 2023 to 2024, he was a Graduate Teaching Assistant with the Department of Physics and Astronomy, West Virginia University, where he has been a Graduate Research Assistant since 2024. His research interests include quantum optics, with a focus on semiconductor quantum dots as quantum emitters, as well as the design, simulation, optimization, and fabrication of nanophotonic devices.

\end{IEEEbiographynophoto}

\begin{IEEEbiographynophoto}{Tyler Manfre}
received the B.S. in Physics from West Chester University in 2025 and the M.S. in Quantum Science and Engineering from the University of Delaware in 2026. From 2025 to 2026 he was a graduate student researcher at the University of Delaware.

His research interests include nanofabrication, semiconductor processing, photonics, and quantum technologies.
\end{IEEEbiographynophoto}

\begin{IEEEbiographynophoto}{Nazifa T. Arony}
received the B.S. degree in physics from the University of Rajshahi, Rajshahi, Bangladesh, in 2017, and the M.S. degree in physics from the same university in 2019. She received the M.S. degree in materials science and nanotechnology from Bilkent University, Ankara, Turkey, in 2020. She is currently pursuing the Ph.D. degree in materials science and engineering at the University of Delaware, Newark, DE.

She has been a Graduate Research Assistant with the Department of Materials Science and Engineering at the University of Delaware since 2021. Her research interests include understanding the properties of patterned quantum dots grown by molecular beam epitaxy.

\end{IEEEbiographynophoto}

\begin{IEEEbiographynophoto}{Joshua M. O. Zide}
received the B.S. degree with Distinction in Materials Science and Engineering from Stanford University in 2002 and completed his Ph.D. in Materials at the University of California, Santa Barbara in 2007.

He is a Professor and Chair in the Materials Science and Engineering Department at the University of Delaware, Newark, DE. His research interests focus primarily on the nanoscale engineering of novel semiconductor and composite electronic materials for energy conversion and (opto) electronic devices. He is an author on over 100 publications and holds several patents.

Prof. Zide has received the International Thermoelectric Society Goldsmid Award (2007), a Young Investigator Award from the Office of Naval Research (2009), the North American Molecular Beam Epitaxy Young Investigator (2011), a Department of Energy Early Career Award (2012), and the AVS Peter Mark Memorial Award (2014) and was named a Fellow of AVS in 2021.

\end{IEEEbiographynophoto}

\begin{IEEEbiographynophoto}{Matthew F. Doty}
(M’11) received the B.S. in Physics from the Pennsylvania State University in 1998, and the M.S. and Ph.D. from the University of California, Santa Barbara in 2001 and 2004, respectively. 

From 2004 to 2007, he was a National Research Council Research Associate at the Naval Research Laboratory Research in Washington, DC. Since 2007, he has been a Professor in the Department of Materials Science and Engineering at the University of Delaware, Newark, DE. He is the founding director of the University of Delaware’s Quantum Science and Engineering Program and the author of more than 90 articles. His research interests include light-matter interactions and the development of new material and device paradigms for future technologies including improved solar energy harvesting, optical control of magnetism, and integrated quantum photonics. Dr. Doty was a 2009 recipient of an NSF Career award.

\end{IEEEbiographynophoto}

\begin{IEEEbiographynophoto}{Edward B. Flagg}
received the B.S. degree in physics from the Massachusetts Institute of Technology in 2001 and the Ph.D. in physics from the University of Texas at Austin in 2008. 

From 2008 to 2012, he was a Research Associate at the National Institute of Standards and Technology in Gaithersburg, MD. Since 2013, he has been a Professor with the Department of Physics and Astronomy at West Virginia University in Morgantown, WV. His research interests include condensed matter quantum optics, nonlinear optics, nanophotonics, and quantum computing.

Prof. Flagg is a member of APS and Optica. He received the NSF CAREER Award in 2015, the Cottrell Scholars Award from the Research Corporation for Science Advancement in 2017, and the West Virginia University Foundation Outstanding Teacher Award in 2018.

\end{IEEEbiographynophoto}

\end{document}